# Journal Name

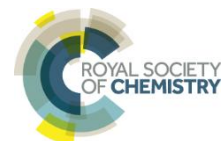

## ARTICLE

# Shearing-induced contact pattern formation in hydrogels sliding in polymer solution

Shintaro Yashima[ab], Satoshi Hirayama[a], Takayuki Kurokawa[cd]*, Thomas Salez[de], Haruna Takefuji[f], Wei Hong[dgh] and Jian Ping Gong[cdi]*

The contact of a hydrogel during the rotational shearing on glass surface in concentrated polymer solution was observed *in situ*. Dynamic contact patterns that rotate in-phase with the rotational shearing of the gel were observed for the first time. The contact patterns with a periodicity in the circumferential direction appeared and became fine with the shearing time. The patterns appeared more quickly at elevated sliding velocity, polymer concentration, and normal pressure. Furthermore, the softness of the gel also substantially influenced the character of the patterns. The pattern formation was discussed in terms of the non-linear rheology of the polymer solution at the rotational soft interface.

## Introduction

Biological systems display fascinating low friction properties, as find in human joints and eyes. The friction events in the biological systems take place between water-containing soft tissues intermediated by viscoelastic synovial fluid or mucus.[1-5] Hydrogels are a class of material similar to biological soft tissues, in terms of their water containing structure and softness. Hydrogels are therefore used as model substances to elucidate the low friction mechanism of soft tissues as well as biomaterials such as artificial cartilages and meniscus.[6-11] Previous researches have shown that hydrogels exhibit very rich friction behaviors, in which solid friction and fluid dissipation coexist. The surface chemistry, surface topology, bulk mechanical property of the hydrogels, and the polymer lubricants are all play roles in intertwined manner.[12-20]

To elucidate the effect of synovial fluid, hydrogel friction in polymer solutions have been studied previously.[14,17] It has been found that the friction at low sliding velocity, at which the interfacial interaction is the dominant mechanism, decreases with the polymer concentration in the dilute and semi-dilute regimes, reaching a minimum value for a concentration around 10C*, and then the friction increases with the polymer concentration in the concentrated region. Here, C* is the overlap concentration of the polymer solution. Such an effect of polymer solution has been observed not only for sodium hyaluronate (HA) that is the main component of synovial fluid but also for poly(ethylene oxide), of various molecular weights, suggesting the generic effect of polymer solution as lubricants.[14,17] The polymer concentration dependence could be associated to two opposite effects to the hydrogel friction. The polymer at the interface screens the direct contact of the hydrogel to the substrate, which reduces the friction. On the other hand, the enhanced viscosity of the lubricant layer increases dissipation. As the viscosity of the polymer solution rapidly increases with the concentration above C*, the viscosity effect overwhelms the screening effect at high concentration. Direct observation of interfacial contact during the friction process is indispensable for elucidating the effect of polymer solution on the friction of hydrogels. In this work, we *in-situ* observed the macroscopic contact at the sliding interface of polyvinyl alcohol (PVA) hydrogels on glass substrates in concentrated HA solutions. The friction experiment was performed using a strain-controlled parallel-plate rheometer. The sliding interface was observed using a recently developed optical system based on the principle of critical refraction. Unique contact patterns with circular periodicity appeared with the sliding time. The effect of polymer concentration, sliding velocity, normal load, and modulus of the gel on the characters of the dynamic patterns were investigated. The probable mechanism for the contact pattern formation is discussed using the concept of gel/glass surface instability.[21-24]

## Methods

### Sample preparation

**Hydrogels:** The physically crosslinked polyvinyl alcohol (PVA) gels were used in this study. The gels were prepared by the quenching method from PVA solution of mixture solvent (dimethyl sulfoxide:$H_2O$=3:1, w/w). Unless specified, the PVA solution of 10 wt% concentration was prepared by heating PVA (Mw=88,000,

[a.] Grad. School of Life Science, Hokkaido University, Sapporo, Japan
[b.] Dept. Chemistry, Fac. of Science, Kyusyu University, Fukuoka, Japan
[c.] Fac. Advanced Life Science, Hokkaido University, Sapporo, Japan
E-mail: gong@sci.hokudai.ac.jp
[d.] Soft Matter GI-CoRE, Hokkaido University, Sapporo, Japan
[e.] Univ. Bordeaux, CNRS, LOMA, UMR, Talence, France
[f.] Dept. Chemistry, Grad. School of Science, Kyushu University, Fukuoka, Japan
[g.] Dept. Aerospace Engineering, Iowa State University, Ames, Iowa, United States
[h.] Dept. Mechanics and Aerospace Engineering, Southern University of Science and Technology, Shenzhen, Guangdong, China
[i.] Inst. Chemical Reaction Design and Discovery (WPI-ICReDD), Hokkaido University, Sapporo, Japan

† Electronic Supplementary Information (ESI) available: Estimation of lubricant film thickness, contact images using water as a lubricant and video files. See DOI: 10.1039/c8sm02428f

Polymerization degree=2,000, Nacalai Tesque Co., Ltd) in the mixture solvent for 1 hour at 90˚C. After heating, the degassed PVA solution was poured into the reaction cell that consists of two glass plates with a silicone rubber spacer (3.0 mm) to form a hydrogel sheet, and this hydrogel sheet was quenched at -40˚C for 16 h. The hydrogel sheet was then allowed to warm up to room temperature and subsequently immersed in a large amount of distilled water for one week to extract the dimethyl sulfoxide. The water content, thickness, and Young's modulus $E$ of the equilibrium-swollen gels were 19.5 wt%, 2.4 mm, and 40 kPa, respectively. The Young's modulus was measured by using a tensile-compressive tester (Tensilon, Orientec., Co.). The disc-shaped sample of 10 mm in diameter and 2.4 mm in thickness was compressed with a velocity of 10% thickness min$^{-1}$. Using the relation of strain and nominal stress, the Young's modulus was calculated at a strain of less than 10%. The modulus data was averaged over 3 samples.

**Polymer solution as a lubricant:** HA with molecular weight of 1.9x10$^6$ g/mol was used. The HA aqueous solutions of various concentrations (10C* ~60C*) containing 0.1 M NaCl to inhibit aggregation of HA and 0.2 g/L sodium azide for suppression of bacterial growth were prepared. In 0.1 M NaCl, the overlap concentration, $C^*$, of HA is 0.260 g/L.[25] In the studied concentration range of 10C* ~ 60C*, the HA concentrations were 0.26 wt% ~ 1.56 wt%. At the C* concentration, the osmotic pressure from the polymer is 1.27 Pa and from the Donnan effect is 2.58 Pa.[25] Accordingly, the overall osmotic pressures of the HA solutions were 31 Pa ~ 230 Pa, which were much lower than the modulus of the PVA gels. This ensures that the PVA gels were not dehydrated in the HA solution, and furthermore, it also ensures a large enough difference in the refractive indices between the gel and the HA solution, which is necessary for the optical observation of the contact interface.

**Glass surfaces:** Hydrophobically treated cover glasses (Micro cover glass, C050701; Matsunami Glass Ind. Ltd.) were used as rigid counter surfaces for the observation during sliding friction. For the surface treatment, these glass pieces were soaked in alkali solution for 1 hour and washed with water. Furthermore, an ozone-cleaning chamber (UV/O$_3$) was used for extra surface cleaning. To prepare the hydrophobic surface, the cleaned glasses were further exposed to binding silane vapor, 1$H$, 1$H$, 2$H$, 2$H$ – Perfluorodecyltrichlorosilane (FDTS), at 20 kPa for 8 hours in a vacuum desiccator. The contact angle of water on the cover glass with FDTS coating was 110 ± 2˚.

**Measurements**
**Friction measurement and in situ observation setup of the gel-glass interface:** The friction measurement and *in-situ* observation of the gel-glass interface were performed at the same time. The friction of hydrogels against glass surfaces was measured using a rheometer (ARES, Rheometric Scientific F. E. Ltd.) that works in a constant compressive strain mode at 25$^o$C. The rheometer, with parallel-plate geometry, was equipped with a home-made optical system, which allows the *in situ* observation of the gel-glass interface (Fig. 1).[26] The equilibrium-swollen gels were cut into a disc shape of $R$=7.5 mm in radius and glued on the lower plate of the rheometer using cyanoacrylate instant adhesive agent (Toa Gosei Co., Ltd.). Small trapezoidal prism was fixed to the upper plate of the rheometer, and a cover glass was fixed to the prism as the counter surface of friction. Hydrogel was bonded to the rotating lower plate of the rheometer. (b) The hydrogel was compressed in the polymer solution by the upper plate with a prism. The images of the frictional interface were recorded *in situ* by using a digital video camera set at an observation angle between the critical angles of polymer solution and gel.

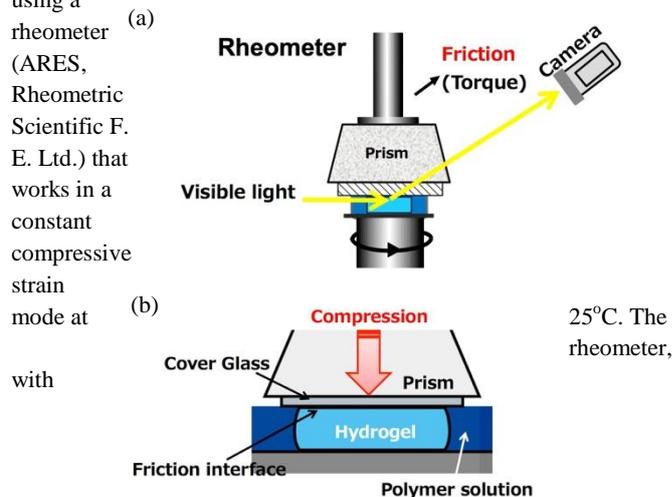

Fig. 1 Schematic illustration of the setup for the *in-situ* observation of the frictional interface between a gel and a glass during sliding friction measurement. (a) A parallel-plate rheometer was used. The amount of polymer aqueous solution of prescribed concentration as a lubricant was poured on the gels to cover their whole surface. As the counter surface, the cover glass treated as described above was attached to the bottom of the prism surface, and the prism was attached to the upper plate of the rheometer. The counter surface was brought towards the gel surface at a constant speed of 2 μm/s until the initial normal force reached the prescribed value. The rotation of the lower plate attached to the gel was started immediately after the normal force reached the prescribed value. The direction of rotation was clockwise in all measurements. The angular velocity $\omega$ of the lower plate was varied in the range of 1.3 rad/s to 13.5 rad/s. The rotation lasted for a prescribed time at a constant angular velocity. The images of the frictional interface were recorded *in situ* by using a digital video camera (HDR CX550, Sony) set at an observation angle that is larger than the critical angle of the polymer solution but smaller than the critical angle of the gel. The video camera was located 1.5 m away so that the whole sample could be observed at the observation angle. In the contact images, the bright regions correspond to the gel in direct contact with the glass substrate, and the dark regions correspond to a gap with polymer lubricants. The friction force, $F$, was calculated as $F = 4T/3R$, where $T$ was the frictional torque recorded during measurement and $R$ (=7.5 mm) was the radius of the apparent contact area.[27] Here, we assumed that friction is linearly proportional to the velocity, and the torque $T$ was the sum over the velocity range of 0 (center of axis) to $\omega R$ (periphery of gel). It should be noted that the "friction" described here refers to all the dissipative force at the gel/glass interface, including i) the dry friction at the gel/glass interface and ii) the viscous resistance of lubricant. Therefore, we represented friction stress the "shear stress at the gel/glass interface" for clarity. The average shear stress, $\sigma$, generated at the gel/glass interface was defined as the friction force per unit area, $\sigma = F/\pi R^2$.

**Viscosity measurement:** The viscosity of the HA solutions was determined using a commercially available rheometer (ARES-G2) with cone-plate geometry of 50 mm in diameter at 25°C, covering a shear rate range from $10^{-3}$ to $10^3$ s$^{-1}$.

# Results

To reveal the rheological behaviors of the polymer solutions used as the lubricant, the shear rate dependency of the shear stress and viscosity of HA at various concentrations (10C* ~ 60C*) was investigated. As shown in Fig. 2a, the shear stress $\tau$ increases with the shear rate $\gamma$ and saturates at high shear rate, indicating shear thinning. For the same shear rate, the shear stress increases with the HA concentration. As shown in Fig. 2b, all the samples exhibit shear thinning, where the viscosity begins to decrease as the shear rate is increased beyond a typical threshold value. With the increase of HA concentration, the threshold decreases, indicating that shear thinning occurs earlier for more concentrated HA solutions. We can extrapolate the rheograms to higher shear rate regions beyond the experimental limitation by using the Casson's equation [28, 29],

$$\sqrt{\tau} = k_0 + k_1\sqrt{\gamma} \quad . \quad (1)$$

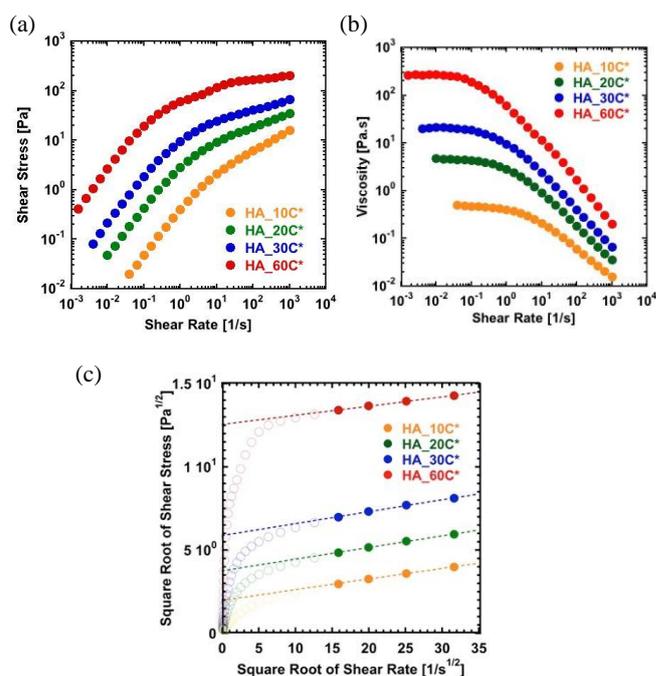

Here $k_0$ and $k_1$ are constants, depending on the polymers and their concentrations. $k_0$ and $k_1$ are determined from the experimentally measured relationship between $\tau$ and $\gamma$ at high shear rate as shown in Fig. 2c. Table 1 shows the $k_0$ and $k_1$ values for HA solutions of various concentrations.

Table 1. $k_0$ and $k_1$ of Casson's equation for HA aqueous solutions (molecular weight 1.9x10$^6$ g/mol) at 25°C.

|  | $k_0$ [Pa$^{1/2}$] | $k_1$ [Pa$^{1/2}$s$^{1/2}$] |
| --- | --- | --- |
| 10C* | 1.967 | 0.064 |
| 20C* | 3.755 | 0.070 |
| 30C* | 5.878 | 0.071 |
| 60C* | 12.55 | 0.055 |

Fig. 2 Dependences of (a) the shear stress and (b) the viscosity, with the shear rate in HA solutions, for various concentrations measured by rheometer at 25°C, as indicated. (c) Casson's plots to determine constants $k_0$ and $k_1$ for HA solutions. C* is the overlap concentration (*i.e.* the onset of the semi-dilute regime). HA molecular weight: 2x10$^6$ g/mol.

Fig. 3a shows the time evolution of the contact pattern at the gel-glass interface for the gel rotationally sliding on the hydrophobic glass in a 30C* HA solution. The images in Fig. 3a were captured from the video in the supplemental information (Video S1). At the very beginning of the sliding, a dark image is observed except at the periphery of the sample, indicating that the HA liquid is entrapped at the interface and the gel has no direct contact with the glass surface except at the edge of the sample. With time, a circular contact pattern with a periodicity in the circumferential direction appears gradually, starting from the periphery and developing towards the center of the sample. The contact pattern rotates in-phase with the rotation of the gel. Together with the contact pattern formation, the total contact area increases. This suggests that the squeezing of lubricants from the interface occurs during sliding. The contact pattern remains even 10 min after stopping the rotation. As shown in Fig. 3b, the corresponding frictional stress increases gradually with time. This increase of friction can be attributed to the increase of contact area during the pattern formation.

the very beginning of shearing ($t \sim 0$) at which the contact is hardly formed (Fig. 3b) is several times higher than the highest shear stress of the HA solution in Fig. 2a, and the average shear rate applied on the intermediated polymer lubricant is much higher than that of the upper limit of the experiments in Fig. 2a, even at the beginning of shearing. Accordingly, we can correlate the frictional stress with the shear rate $\gamma$ through the Casson's equation (1), and the lubricant film

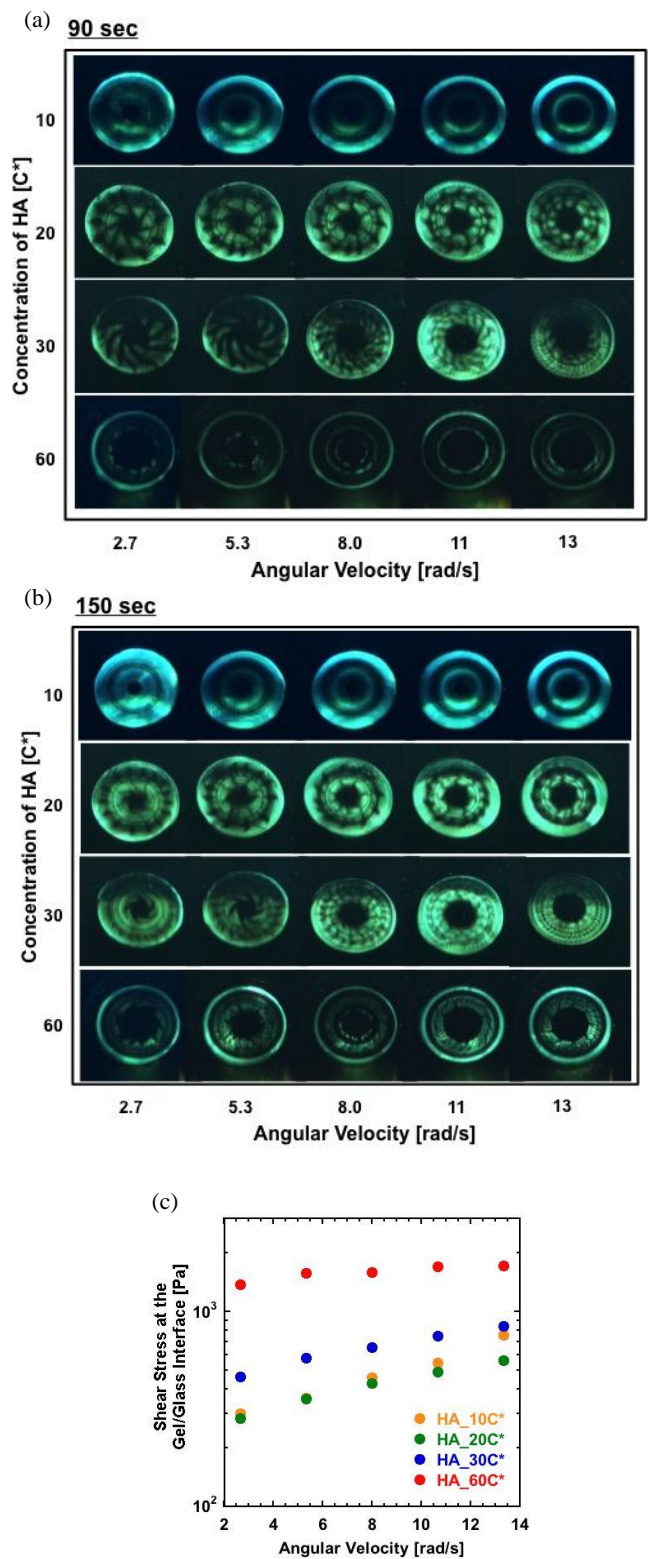

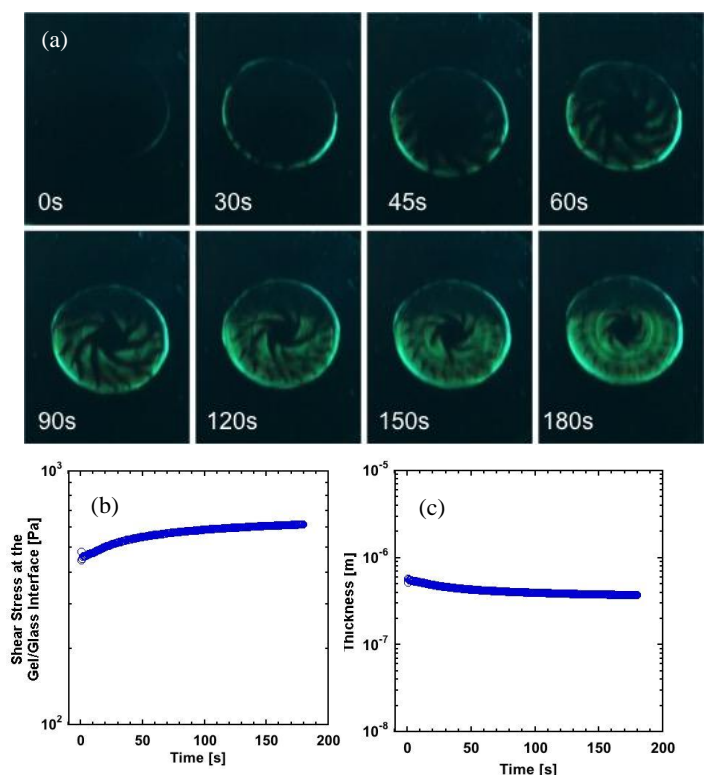

Fig.3 (a) Time evolution of the contact pattern at the contact of the PVA hydrogel/glass interface, together with the corresponding time evolution of (b) the shear stress at the gel/glass interface, and (c) the apparent lubricant layer thickness. The number in each image indicates the elapsed time. Sliding condition: Gel modulus: 40 kPa; lubricant: 30C* HA; normal pressure: 22 kPa; angular velocity: 5.3 rad/s; diameter of the disc: 15 mm.

As shown in Fig. 3a, initially there is no direct contact of the gel on the substrate, and the gel is thus lifted up by the viscous lubrication of the HA polymer solution. The gel/glass interface is the apparent shear rate of the fluid sub-layer can be estimated very roughly, as the ratio of the sliding velocity, $U=\omega r$, to the thickness, $h$, of the fluid film from the measured frictional stress in Fig. 3b, using the rheograms in Fig. 2a. The frictional stress $\sigma$ at

Fig.4 (a, b) HA concentration and angular velocity effects on the contact pattern at the hydrogel/glass interface, and (c) the corresponding shear stresses. The results are for (a) 90 seconds and (b) 150 seconds after the beginning of the sliding motion. Gel modulus: 40 kPa; normal pressure: 22 kPa.

thickness $h$ is related to $\gamma$ as,

$$\gamma = \omega r/h \ , \qquad (2)$$

which varies with the distance $r$ from the center of the sample due to the rotational shearing. $h$ should be a function of time and position. Assuming that $h$ is independent of the distance $r$ from the center, and that the frictional stress all comes from the viscous drag of the lubricant, we estimated the apparent lubricant film thickness ($h$) at different time, as shown in Fig. 3c (Supporting Information). The estimated $h$ at the beginning of sliding is about 1 μm.

The contact pattern of circular periodicity is only observed when the viscous HA solution is used as a lubricant. When water is used as a lubricant, the entire surface of the gel contacts with the substrate immediately, without showing any regular contact patterns even after certain sliding times (Supporting Information, Fig. S1). The frictional stress in water is much higher than that in the HA solution due to direct interaction of the PVA gel with the glass substrate.[17]

The character and dynamics of the contact pattern strongly depends on the concentration of the HA solution and the sliding velocity. Fig. 4a shows a table of images of contact patterns for various HA concentration and angular velocities observed 90 seconds after the starting of the sliding motion. For a 10C* concentration, only two thick concentric circles are observed, and this pattern becomes clearer at high angular velocity. For a 20C* concentration, two concentric circles superimposed with periodic patterns in the circumferential direction appear. The circumferential pattern becomes finer with smaller domain size, but the concentric circles become less clear at high velocity. With the further increase of HA concentration to 30C*, the two concentric circles merge together and the circumferential pattern develops into even finer domain sizes. For 60C*, only two thin concentric circles with a circumferential pattern are observed. Apparently, the squeezing of the liquid from the interface is too slow to form the contact at the given observation time (90 seconds), due to the very high viscosity of the 60C* solution. The table of images of contact patterns with varied HA concentrations, and angular velocities observed at 150 seconds is shown in Fig. 4b. Essentially, Fig. 4b shows similar patterns as those observed in Fig. 4a for 90 seconds but the patterns display finer characters, especially for 60C*. Each contact pattern becomes finer with the increase of the angular velocity when compared at the same sliding time. As the concentration increases, that is, as the viscosity of the lubricant increases, the period in the circumferential direction tends to become smaller. Comparing Fig. 4a and Fig. 4b, it is apparent that the total contact area increases with the elapsed time, which also indicates the squeezing of lubricants from the interface as seen from Fig. 3a. The squeezing effect is more prominent for the low HA concentration and becomes less remarkable for high HA concentrations, which can be attributed to the increase of viscosity at high HA concentration.

Fig. 4c shows the friction behavior corresponding to the gel/glass interface images shown in Fig. 4a. The frictional stress increases weakly with the angular velocity for a fixed polymer concentration, which is a result of the coexistence of the solid friction generated at the contact region and the fluid resistance generated at the non-contact region. The slightly higher friction in 10C* concentration than that in 20C* at high velocity suggests that the solid friction dominates in the case of 10C*. The increase in the frictional stress with the polymer concentration from 20C* to 60C* indicates that the viscous drag of the polymer solution dominates the frictional stress at the observed conditions.

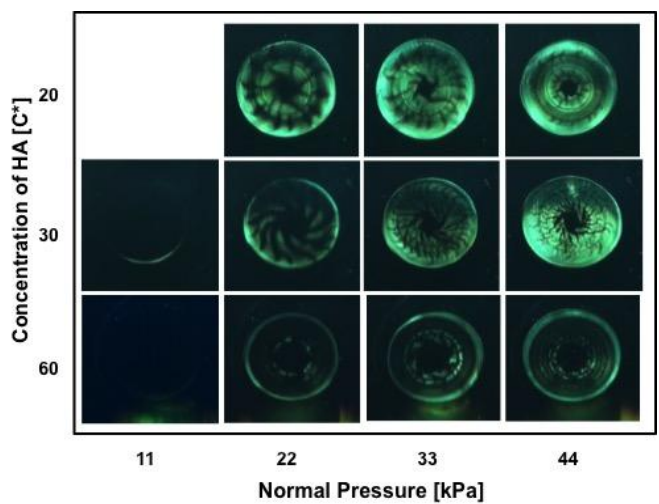

Fig. 5 HA concentration and normal pressure effects on the contact pattern at the hydrogel/glass interface, observed 90 seconds after the beginning of the sliding motion. Gel modulus: 40 kPa; angular velocity: 5.3 rad/s.

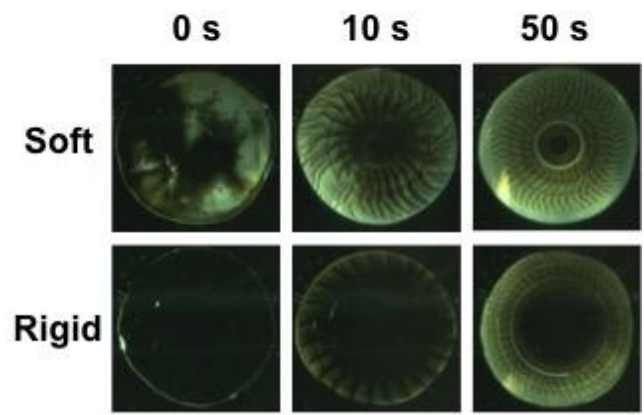

Fig. 6 Time evolution of the contact pattern for soft and hard gels. PVA gels' shear moduli $G'$: 9 kPa (7wt%), 100 kPa (13wt%); lubricant: 30C* HA; normal pressure: 33 kPa; angular velocity: 8 rad/s.

These results are consistent with Fig. 4a that shows large and homogeneous contact areas in the circular direction in the 10C* lubricant while small and patterned contact areas are observed in the 20C*, 30C*, and 60 C* lubricants. As shown in Fig. 4a, in the case of 60 C *, the gel is barely in contact with the glass substrate as compared with other concentrations. The frictional stress must come from the viscous resistance of the fluid. However, the frictional stress has weak velocity dependence as shown in Fig. 4c, one can deduce that strong shear thinning occurs in the 60C* system.

To investigate the squeezing effect during sliding, the experiment was further performed at various normal pressures. Fig. 5 is the table of images showing the effect of normal pressure at an angular velocity of 5.3 rad/s and observed at 90 seconds. Patterns appear only at relatively large normal pressures (> 22 kPa) at the given observation time, and the patterns become finer with a smaller domain size when increasing the normal pressure. The result indicates that high normal pressure accelerates the squeezing of lubricant from the interface, and therefore, accelerates the contact pattern formation.

To study how the elasticity of the gels plays a role, we further performed the experiment using two gels of quite different moduli. Fig. 6 shows the time evolution of the contact images for soft and hard gels with Young's moduli of 27 kPa and 300 kPa, respectively. Under the same normal pressure (33 kPa), large part of the soft gel already formed contact, while the hard gel only had some contact at the periphery.   For the soft gel, applying a shear rotation immediately induces contact patterns in all the pre-contacted area. While it took a longer time for the hard gel to develop the contact pattern towards the inner region. In addition, the contact patterns of the soft gel exhibit larger sizes with characters strongly twisted by the shear stress imposed on the soft gel. These results clearly show that with a soft gel, the contact pattern forms more quickly, displaying clear periodical characters.

## Discussion

The dynamic contact pattern formation indicates the occurrence of an instability at the interface. Shearing is found to be essential for this instability, as we do not observe any pattern at a constant normal compression without any shear. This instability is not related to the Schallamach wave, in which the pattern travels in the opposite direction with the shearing, together with a stick-slip motion.[30] In the present case, the shearing and pattern rotations are in phase and there is no stick-slip like motion. Also, the shear velocity is too low for inertia to play any role. To see if the instability is governed by the local pressure variations, we compare the ratio of the frictional stress $\sigma$ to the normal pressure $P$, $\sigma/P$, to the ratio of the lubricant film thickness $h$ to the wavelength of the pattern $\lambda$, $h/\lambda$, according to the Stokes equation in the lubrication approximation.[31] Reading from Fig. 3 and Fig. 4, $\sigma/P$ is on the order of 0.1-0.01. The thickness $h$ is roughly estimated as ~1 μm from the Casson's plot, following the literature (see supporting information), while $\lambda$ reaches several millimeters, which gives $h/\lambda$ less than 0.001. This means that the patterns we observed are at least one order of magnitude larger than expected from the pressure variations.

Based on the experimental observations, we believe that there are two types of (interrelated) instabilities in the system, both involving the non-Newtonian rheology of the HA fluids. The primary mode of instability appears as annular contact patterns, as seen in the cases of low HA concentrations (Fig. 4a), and is believed to be a direct consequence of the shear thinning rheology of the fluid. For a non-Newtonian fluid with a rheology similar to the Casson's model, there exists a critical shear rate (or a yield stress) beyond which shear thinning occurs. The rotation in the rheometer provides a large enough circumferential shear such that the fluid in most regions (except the center) is in the shear thinning regime. In other words, the rotation yields the fluid, and enables significant gap flow in the radial direction, driven by the relatively small pressure gradient. Moreover, under a constant gap thickness, the circumferential shear rate is proportional to the distance to the rotation center, and thus the viscosity of the shear thinning fluid is smaller at the outer perimeter. The resulting higher radial flow rate near the outer perimeter tends to deplete the HA solution in the intermediate-radius region while leaving a pocket of fluid in the center, leading to annular contact patterns.   With the driving force for the radial flow being the radial pressure gradient $\nabla p$ and the resistance being related to the viscosity of the HA solution, the process has two dimensionless parameters, $\nabla p H/k_0^2$ and $\nabla p H/k_1^2 \omega$, where $H$ is the initial thickness of the fluid, and $\omega$ the angular rotation velocity. The pressure gradient buildup is related to those two dimensionless parameters, e.g. it is expected that solutions of higher $k_0$ (higher HA concentration) and/or rotating at a higher angular velocity tend to exhibit larger pressure gradients.

The secondary instability generates the radial-like channels through the annular contact region. The process is similar to the gap flow between a soft material initially in contact with a rigid wall: instead of uniformly opening the gap, the fluid tends to tunnel through the interface by locally deforming the soft solid. The characteristic length (circumferential wavelength) $\lambda$ of the instability results from the competition between the shear elastic modulus $\mu$ of the soft solid and the radial pressure gradient:

$$\lambda \sim \frac{\mu}{\nabla p} \quad . \tag{3}$$

Comparing to that caused by rotation, the shear rate induced by the radial flow is relatively small and thus the effect of shear thinning is minor in this secondary mode of instability. Nevertheless, the characteristic length $\lambda$ is indirectly related to the rheology of the fluid through the pressure gradient $\nabla p$. For example, solutions of higher HA concentrations, or under higher angular velocities, will lead to smaller feature sizes in the radial channels of the contact patterns. These trends agree qualitatively with the experiments, while the detailed model of the process will be addressed in a separate article due to the mathematical complexity of this highly nonlinear problem.

## Conclusions

Dynamic contact patterns displaying periodicity in the circumferential direction are observed during the rotational sliding friction of hydrogels on solid substrate in concentrated polymer solutions. These patterns are formed in the region where the hydrogel and the glass are in contact in the polymer solution under shear stress. Although it belongs to the broad class of pattern formations under shear, it differs from phenomena such as shear banding in that the polymer solution itself forms a pattern. The appearing time and character of the contact patterns are dependent on the polymer concentration, the sliding velocity, the normal pressure, and the modulus of the gel. More concentrated polymer solutions and higher sliding velocities produce contact patterns with finer characters. Higher pressures and lower moduli lead to a quicker pattern formation. The results suggest that the instability is related to the shear thinning of the HA solutions as lubricants of the soft hydrogels under rotational shearing.

## Conflicts of interest

There are no conflicts to declare.

## Acknowledgements

This work was supported by JSPS KAKENHI Grant Numbers JP24120003, JP17H06144, and JP17H06376. Institute for Chemical Reaction Design and Discovery (ICReDD) was established by World Premier International Research Initiative (WPI), MEXT, Japan.

# Supporting Information

**Shearing-induced contact pattern formation in hydrogels sliding in polymer solution**

*Shintaro Yashima, Satoshi Hirayama, Takayuki Kurokawa\*, Thomas Salez, Haruna Takefuji,*

  *Wei Hong, Jian Ping Gong\**

**Estimation of lubricant film thickness *h***

We assume that lubricant film thickness $h$ is independent of the distance $r$ from the center of the gel disc, $h(r)=h$. The frictional force $F$ is given by:

$$F = \int_0^R \tau \, 2\pi r \, dr \quad, \tag{S1}$$

where $\tau$ is the shear stress and $R$ is the radius of the gel disc.

The shear rate $\gamma$ is written as:

$$\gamma = \frac{\omega r}{h} \quad. \tag{S2}$$

Assuming the shear rate $\gamma$ to be high, by inserting Casson's equation (1) in equation (S1), the frictional force can be expressed as:

$$F = \int_0^R (k_0 + k_1\sqrt{\gamma})^2 2\pi r \, dr = 2\pi \int_0^R \left(k_0^2 + 2k_0 k_1 \sqrt{\frac{\omega r}{h}} + k_1^2 \frac{\omega r}{h}\right) r \, dr$$

$$= 2\pi \left[\frac{1}{2} k_0^2 R^2 + \frac{4}{5} k_0 k_1 \sqrt{\frac{\omega}{h}} R^{\frac{5}{2}} + \frac{1}{3} k_1^2 \frac{\omega}{h} R^3\right] \quad. \tag{S3}$$

The friction stress $\sigma = F/\pi R^2$ can be written as:

$$\sigma = k_0^2 + \frac{8}{5} k_0 k_1 \sqrt{\frac{\omega R}{h}} + \frac{2}{3} k_1^2 \frac{\omega R}{h} \quad. \tag{S4}$$

Rewriting this quadratic equation (S4) by setting $x=h^{-1/2}$, one gets:

$$\frac{2}{3} k_1^2 \omega R x^2 + \frac{8}{5} k_0 k_1 \sqrt{\omega R} \, x + k_0^2 - \sigma = 0 \quad. \tag{S5}$$

The lubricant film thickness $h$ is then given as the solution of the latter:

$$h = \frac{1}{x^2} = \frac{k_1^2 \omega R}{\left(-\frac{6}{5} k_0 + \sqrt{\frac{3}{2}\sigma - \frac{3}{50} k_0^2}\right)^2} \quad (\because x > 0) \quad. \tag{S6}$$

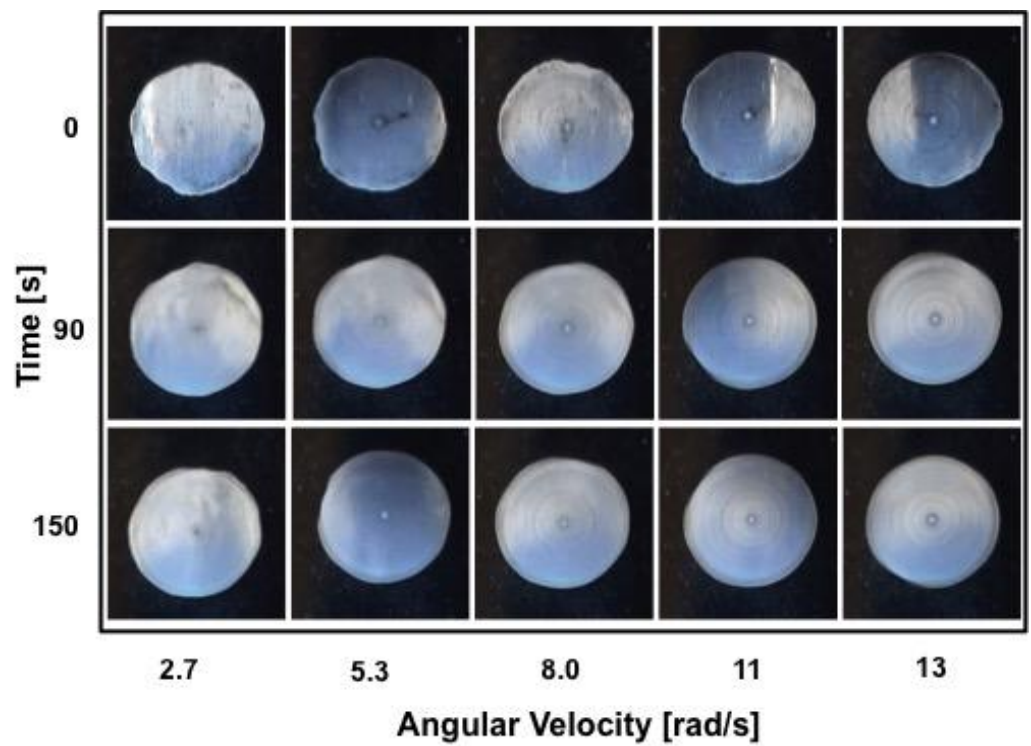

Fig. S1 Time and angular velocity effects on the contact at hydrogel/glass interface when water was used as a lubricant. Gel modulus: 40 kPa; normal pressure: 22 kPa.

Video S1 Video image of contact pattern at hydrogel/glass interface under sliding motion at angular velocity 5.3 rad/s.　Gel modulus: 40 kPa; lubricant: 30C* HA; normal pressure: 22 kPa. The images shown in Fig. 3 were extracted from the Video S1.

Video S2 Video image of contact pattern at hydrogel/glass interface under sliding motion at angular velocity 13 rad/s. Gel modulus: 40 kPa; lubricant: 30C* HA; normal pressure: 22 kPa. The images shown in Fig. 4a and Fig. 4b were extracted from the Video S2.